# Side Channel Attacks on STTRAM and Low-Overhead Countermeasures

Nitin Rathi, *Helia Naeimi and Swaroop Ghosh
Computer Science and Engineering, University of South Florida
*Intel Labs, Santa Clara, CA
{nitinr, swaroopghosh}@mail.usf.edu, helia.naeimi@intel.com

***Abstract- Spin Transfer Torque RAM (STTRAM) is a promising candidate for Last Level Cache (LLC) due to high endurance, high density and low leakage. One of the major disadvantages of STTRAM is high write latency and write current. Additionally, the latency and current depends on the polarity of the data being written. These features introduce major security vulnerabilities and expose the cache memory to side channel attacks. In this paper we propose a novel side channel attack model where the adversary can monitor the supply current of the memory array to partially identify the sensitive cache data that is being read or written. We propose several low cost solutions such as short retention STTRAM, 1-bit parity, multi-bit random write and constant current write driver to mitigate the attack. 1-bit parity reduces the number of distinct write current states by 30% for 32-bit word and the current signature is further obfuscated by multi-bit random writes. The constant current write makes it more challenging for the attacker to extract the entire word using a single supply current signature.***

## I. Introduction

Spin-Transfer Torque RAM (STTRAM) [1] is promising for Last Level Cache (LLC) due to numerous benefits such as high-density, non-volatility, high-speed, low-power and CMOS compatibility. Fig. 1 shows the STTRAM cell schematic with Magnetic Tunnel Junction (MTJ) as the storage element. The MTJ contains a free and a pinned magnetic layer. The resistance of the MTJ stack is high (low) if free layer magnetic orientation is anti-parallel (parallel) compared to the fixed layer. The MTJ can be toggled from parallel to anti-parallel (or vice versa) by injecting current from source-line to bitline (or vice versa). The data in MTJ is stored in the form of magnetization. The data stored is '1' if the free layer magnetization is anti-parallel to fixed layer magnetization and '0' if they are parallel. The read/write latency of MTJ depends on the size of the device, current passing through the layers as well as on process variation.

STTRAM depends on ambient parameters like magnetic field and temperature that can be exploited to tamper with the stored data. The free layer of MTJ flips under the influence of external magnetic field which can be exploited by the adversary to launch magnetic attacks using a horseshoe magnet or an electromagnet [2]. The switching of MTJ depends on the ambient temperature, at high temperature the MTJ resistance reduces resulting in high read and write current [3]. The increased read current leads to read disturb failures, where the bits are accidentally flipped during read operation because the read current becomes higher than the critical current. The temperature can also be exploited to extend the retention time of the memory [11]. At lower temperature the retention time increases providing more time for the adversary to launch attacks in volatile and semi-nonvolatile memories. The persistent user data in non-volatile cache can also be compromised by launching intentional read and write operation and probing the data buses after the authentic user has logged off. The persistent data leaving the cache can also be accessed by probing the data buses connecting the cache and CPU and the cache and main memory [4].

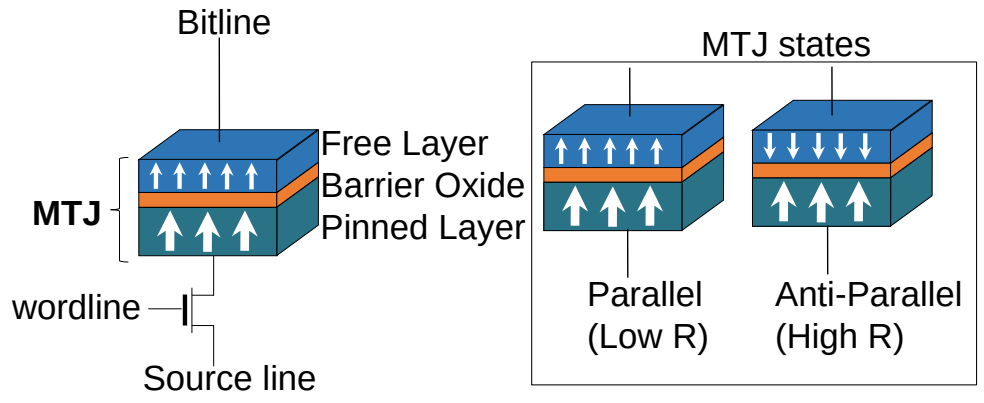

Fig. 1 Schematic of STTRAM bitcell showing MTJ.

Traditional attacks can also be extended for STTRAM such as, (a) micro-probing, where conductors are attached to the chip surface directly to interfere with the integrated circuit; (b) radiation imprinting, where the contents are burned in using

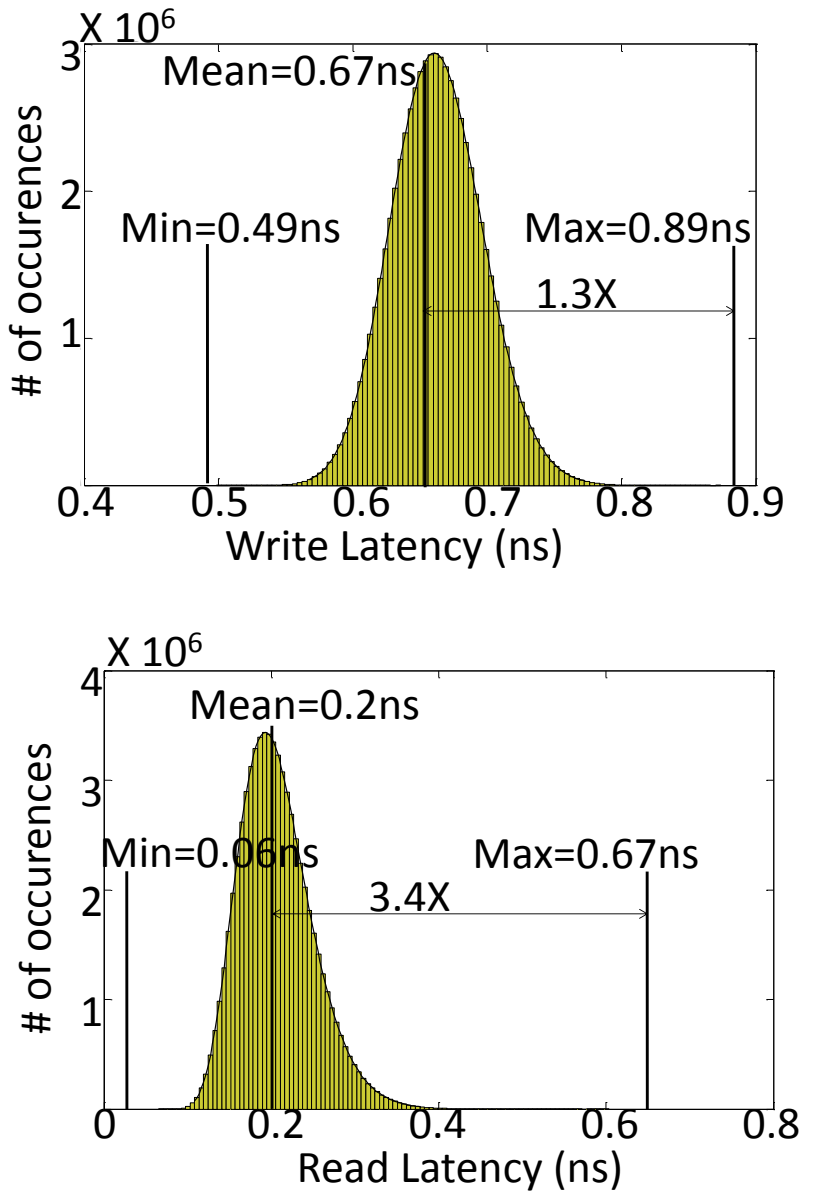

(a)

(b)

Fig. 2 (a) Write latency; (b) read latency distribution of an 8MB STTRAM cache under process variation. The long read and write latency presents wider attack window to the adversary.

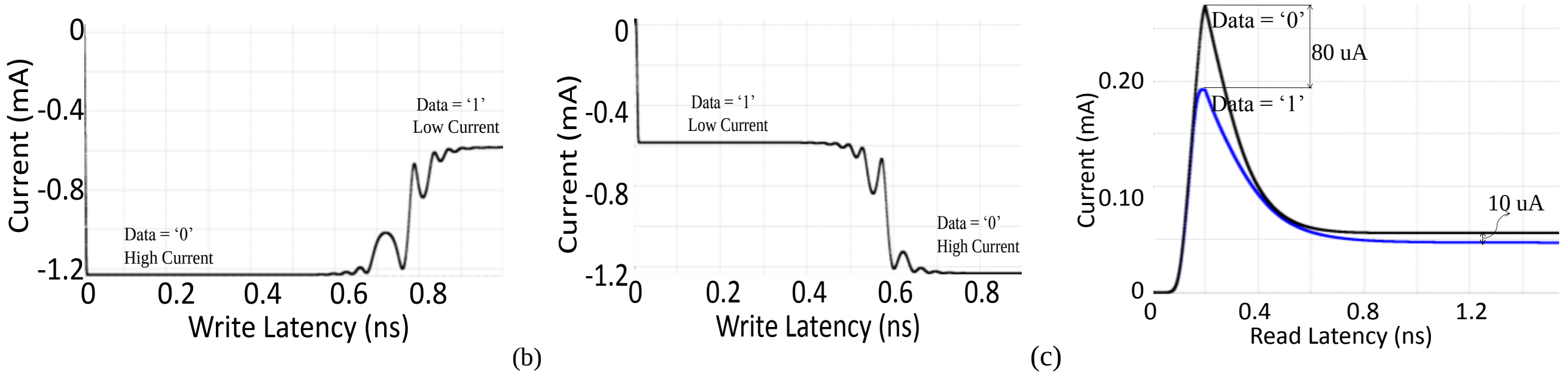

Fig. 3 (a) Supply current waveform for write '1'; (b) write '0', and (c) read operations. A significant gap is present between write '0' and '1' as well as read '0' and '1' currents which can be employed as signature. Furthermore, the magnitude of write current is a function of stored data which also acts as a signature.

X-Ray radiation to prevent overwriting or erasing of stored data; (c) optical probing, where a laser is shinned on the surface resulting in activating the underlying circuit. The active components glow which can then be used to interpret the stored data.

Simple Power Analysis (SPA) is a technique that interprets the measured trace of the supply current to obtain information about the ongoing operation. The current in a circuit can be measured by inserting a small resistance in series with the Vdd or ground rail and then measuring the voltage difference (IR) across the resistance which is converted to instantaneous current. Sophisticated devices can be used to sample the voltage difference at high rates (1GHz) with excellent accuracy (< 1% error) [5]. In SPA the adversary uses the raw measured current or power information to determine the stored data, whereas in differential power analysis (DPA) the adversary uses many measurements to filter out noise. While SPA exploits the relationship between the executed operation and the power leakage, DPA exploits the relationship between the processed data and the power leakage [12]. In this work we focus on both SPA and DPA to launch side channel power analysis attack to obtain the stored data from STTRAM.

We note the fact that STTRAM is associated with high write latency and write current. Furthermore, the write current is asymmetric (polarity dependent). These features introduce major security vulnerabilities as the adversary can monitor these signatures through side channel to compromise data privacy. This is especially possible when crypto processors employ non-volatile memory (NVM) such as STTRAM to store temporary data or the sensitive data is present in the NVM cache during operation in raw form. We also propose low-overhead techniques to obfuscate the side channel signature such as parity encoding, short retention NVM and constant current write. *To the best of our knowledge this is the first effort on STTRAM side-channel attack and preventive techniques.*

In particular, we make the following contributions in this paper. We propose:

(i) Novel vulnerabilities such as long write latency, high write current and asymmetric read/write currents.

(ii) Novel side channel attack models to weaken the data privacy.

(iii) Novel design techniques such as short retention STTRAM, parity encoding and random write to obfuscate the side channel signature

(iv) Constant current write technique to eliminate polarity dependent write current to obfuscate side channel signature.

The rest of the paper is organized as follows. In Section II, we describe the STTRAM vulnerabilities. The attack model is presented in Section III. The preventive measures are described in Section IV. Conclusions are drawn in Section V.

## II. STTRAM Vulnerabilities

In this section we discuss STTRAM vulnerabilities such as high latency, high switching current and asymmetric read/write current.

### *A. Read/Write Latency*

The write latency of STTRAM is a function of thermal stability factor ($\Delta_t$) which in turn depends on the retention time. For 10 year retention $\Delta_t$ =40 is required [13] which corresponds to a write latency of 0.59ns at 1V supply. Furthermore, STTRAM is susceptible to process variation (PV) [6] which increases the thermal stability of bits randomly especially for larger arrays. Therefore, some bits suffer from excessive high read and write latencies. Fig. 2(a-b) shows the read and write latency distribution of a 40nmx40nmx4nm STTRAM under PV. A 5000 point Monte Carlo simulation is performed and the data is extrapolated to 8MB using extreme value theory in Matlab. It is observed that the worst case write (read) latency is 1.3X (3.4X) the mean value. To avoid read and write failures worst case latency is followed for the entire memory array which results in longer wordline pulse. The longer read and write latency presents more opportunity to the adversary to analyze the side channels and weaken the data privacy (Section III).

### *B. Read/Write Current*

Another aspect of STTRAM is the high write current which is dependent on thermal stability, retention time and the polarity of the stored data. We assume constant voltage write which is commonly employed to simplify the write driver design [6]. STTRAM resistance is high (low) during state '1' ('0'). Fig. 3(a) shows the supply current waveform for single bit write '1' when the previous value stored is '0'. Intially the current is high (STTRAM resistance low) and it goes low after successful write (y-axis values are negative). Fig. 3(b) shows the supply current waveform for write '0' with previous value stored as '1', in this case the current is initially low and goes high after successful write. The high and low states of current are very distinct and they reveal the information about the previous and new data. The current difference between the states depends on the Tunnel Magneto Resistance (TMR) of STTRAM which is given by ($R_H$-$R_L$)/$R_L$. For robust read operation it is desired to have higher TMR which adversely affects the data privacy. The read current is comparatively less than the write current (Fig. 3(c)), thus the read and write operation can be distinctly identified from the current waveforms. The source degeneartion based read sensing is used in this work [7].

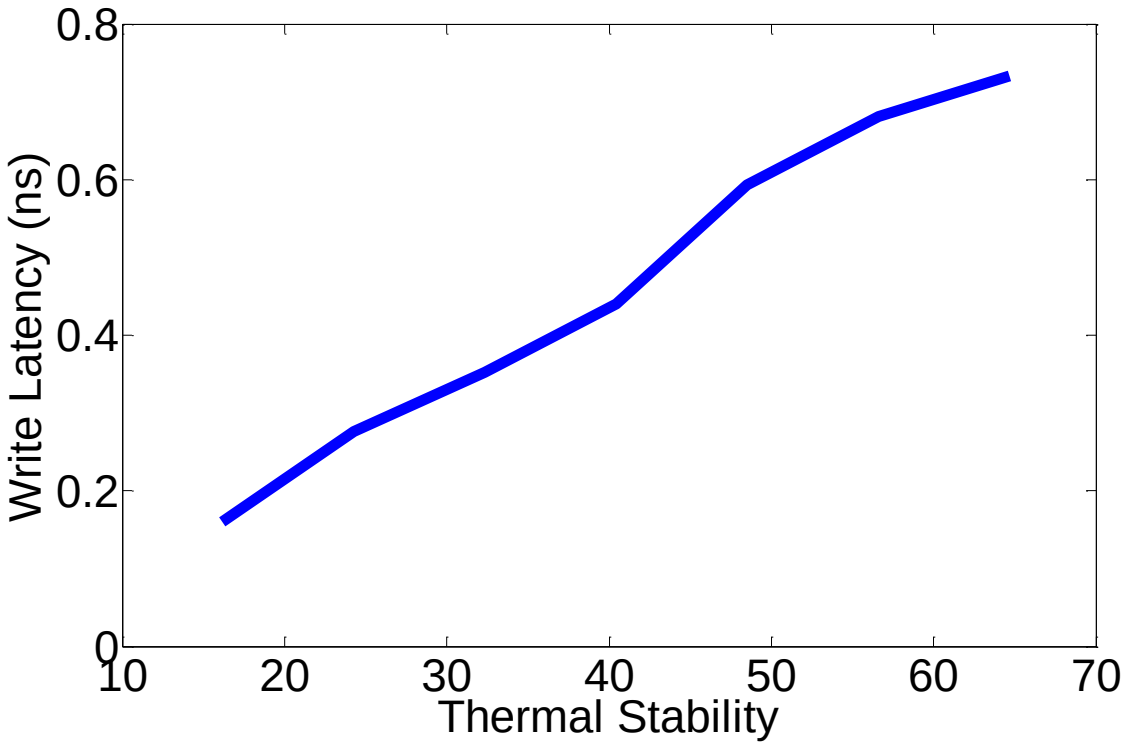

Fig. 4 Write latency for different values of thermal barrier.

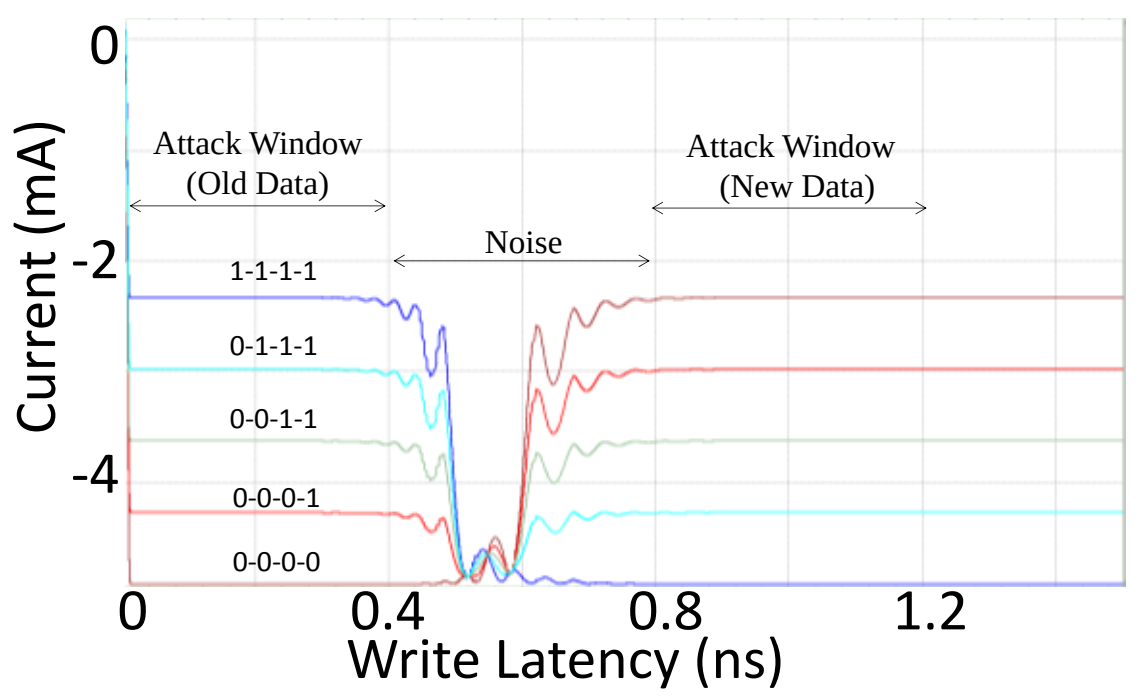

Fig. 5 Write current for 4-bit write operation

### *C. Temperature Sensitivity*

The thermal stability ($\Delta_t$) of STTRAM is a function of ambient temperature and the write current and write latency linearly depends on the thermal stability. The thermal stability is given by $\Delta_t = \frac{H_k M_s A_r t}{2 k_B T}$. *where* $H_k$ *= uniaxial anisotropy,* $M_s$*= saturation magnetization,* $A_r$*= area of MTJ, t=thickness of free layer,* $k_B$*= Boltzmann constant, T= ambient temperature.*

Colder temperature increases the thermal barrier which in turn increases the write current and latency. Fig. 4 shows the write latency for different delta values. The write latency increases with the increase in thermal barrier. This can be exploited by the adversary to strengthen the side channel signature from STTRAM.

## III. STTRAM Attack Models

In this Section we present attack models that builds upon the vulnerabilities described in the previous section.

### *A. Exploiting Read/Write Current*

The LLC contains sensitive data in raw form such as login, password and credit card details entered during a web transaction and encryption keys used to encrypt data to be sent over the network. In current processor architecture all the user data processed by CPU passes through cache memory. The adversary can steal the raw data or get clues about the data so that he can predict the correct data in linear time. For STTRAM LLC the adversary can perform side channel attack by monitoring the supply current waveform of the memory array. It is assumed that the adversary can monitor the current flowing into the memory array from the power supply. Even if the adversary has access to processor power supply, it can reveal the LLC side channel signature. Fig. 5 shows the write current waveforms for 4-bit write operation in STTRAM. Out of 16 data values only 5 are unique in terms of total number of 0's and 1's (1111, 0111, 0011, 0001, 0000). In memory array all the bits in a word are written in parallel, thus the order of 0's and 1's in a word does not affect the supply current waveform rather the overall number of 0's and 1's in a word defines the current signature. For 4 bits all 5 permutations are clearly distinct in the current waveform. Knowing the number of 0's and 1's weakens the security significantly as it reduces the reverse engineering effort to identify the correct data.

### *B. Exploiting Read/Write Latency*

The high read and write latency provides a larger attack window to the adversary. By monitoring the current waveforms the adversary can not only predict the number of 0's and 1's in the new data that is being written but can also predict the previous data by sampling the current just after the wordline is asserted. The adversary samples the current during the attack window shown in Fig. 5. The difference in current states of each combination depends on the TMR of STTRAM as discussed before, higher the TMR more apart are the current states. In Fig. 5 the write operation is completed in 800ps but to avoid write failures under PV the wordline is active for longer duration. This gives adversary more time to identify the transient current and become more confident about the results. Thus, data dependency of current reveals information about the stored and new data and higher latency facilitates the attack. The figure also shows the attack window available to identify the old and new data. Note that larger word size creates more number of states in supply current signature however the difference between two consecutive states remain the same. Furthermore, larger word size increases the total current which makes the attack easier for the adversary.

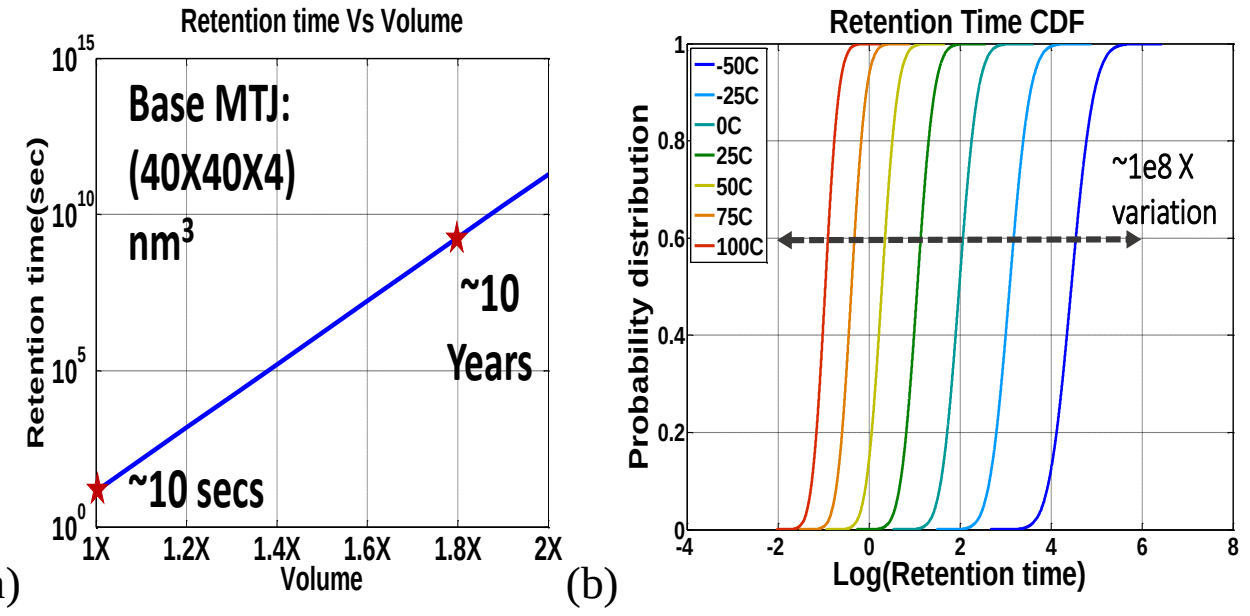

(a) (b)

Fig. 6(a) Retention time variation with respect to MTJ volume; and, (b) retention time dependence on temperature.

### *C. Temperature-Assisted Attack*

The adversary can intentionally increase the write latency by lowering the ambient temperature. The MTJ resistance increases at lower temperatures which leads to less write current. The write latency is directly proportional to the write current and thus at lower temperatures the write latency increases which provides adversary more time to launch the attack.

## IV. Prevention Techniques

In this section we discuss preventive techniques to obfuscate the current signature and/or make the attack difficult or nearly impossible. Since the supply current signature is prominent during write operation we focus our efforts to obfuscate the write current signature.

### *A. Semi Non Volatile Memory (SNVM)*

SNVM is a non-volatile memory with lower retention time. The typical retention time for STTRAM is 10 years however such high retention time is not required for cache application as the data is invalidated when the system restarts or the virtual address space is changed. Instead the retention time can be lowered to improve the write latency and write current [8]. The write latency and write current (I) linearly depends on the thermal barrier ($\Delta_t$) of STTRAM. The retention time (t) is exponentially related to $\Delta_t$ by $t = C \times e^{k\Delta t}$, where C and k are fitting constants.

Both write latency and write current can be lowered by reducing $\Delta_t$ which in turn lowers the retention time. Since $\Delta_t$ depends on the free layer volume of STTRAM it can be scaled to lower the retention time (Fig. 6 (a)). The lower write latency due to SNVM reduces the attack window as shown in Fig. 4. Lower write current brings the current states closer to each other making it difficult to identify the state individually. However, simulations (Fig. 6 (b)) show that at low temperature the retention time increases dramatically, thus giving away the above benefits obtained from lower retention. Thus, SNVM cannot be used in isolation to prevent side channel attack.

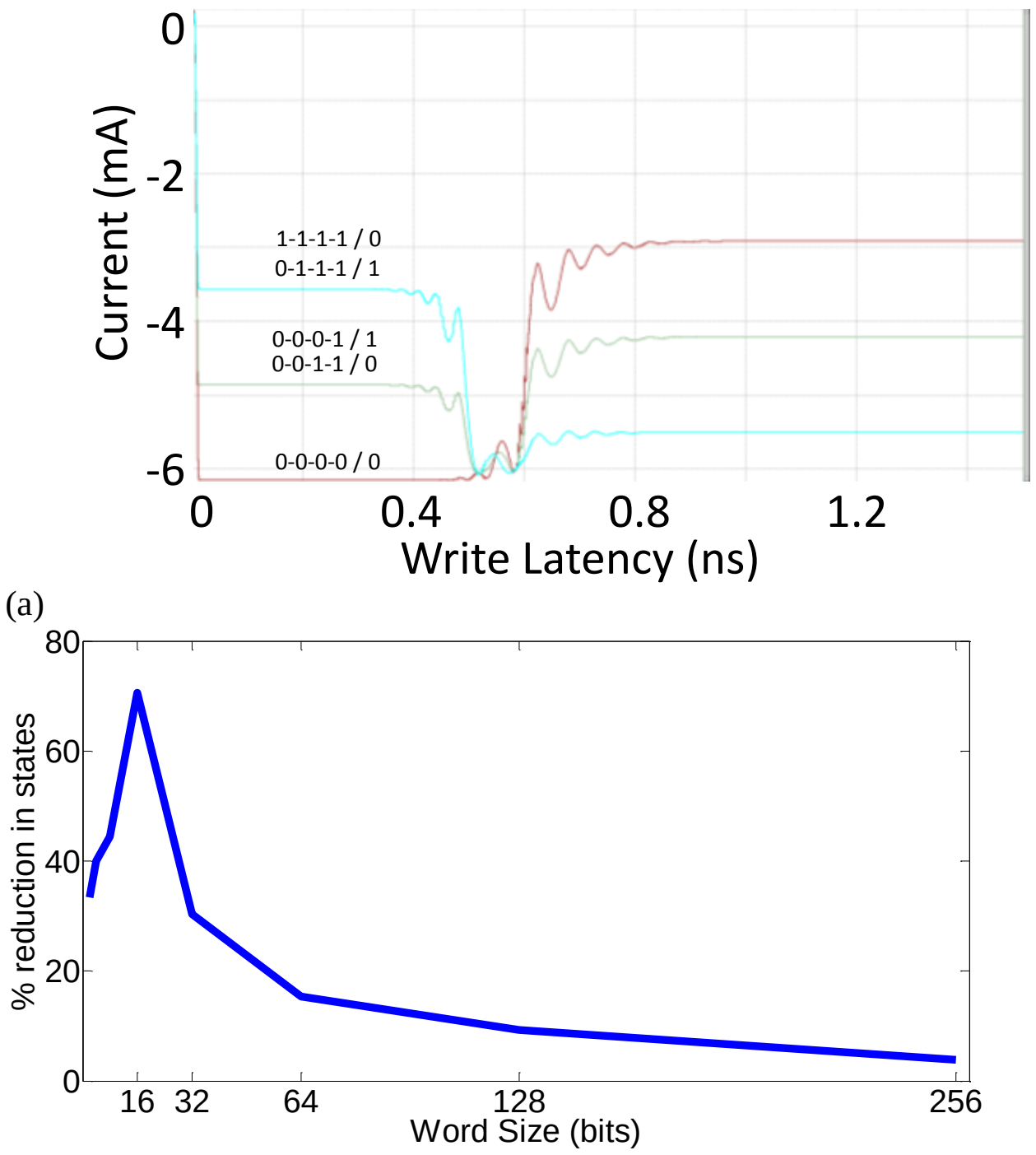

(b)

Fig. 7(a) Current waveform for 4-bit write with 1-bit parity, (b) percent reduction in states with 1-bit parity for different word sizes. Substantial reduction in states in possible with 1-bit parity.

### *B. Adding 1-Bit Parity*

The objective of this prevention technique is to merge multiple supply current levels in the side channel current waveform which will make it difficult for the adversary to predict the states accurately. This is achieved by writing an extra parity bit along with the original data. Fig. 7(a) shows the current waveform of 4-bit write with 1-bit even parity. So, instead of writing 4 bits we write 5 bits with the last bit value decided by the parity of the 4 bits. By doing this we are able to merge 5 states (Fig. 7(a)) into 3 states. Compared to un-coded data the reverse engineering effort increases because a data will map to more number of possibilities. The solution works on the principle that the overall write current depends on the number of 0's and 1's and not on their order. This extra 1-bit write makes some states identical to each other in terms of total 1's and 0's. For example, the un-coded 0111 will become 01111 which will merge with 1111. Fig. 7(b) shows the percent reduction in states with 1-bit parity for different word sizes. For a 32-bit word the number of states reduce by 30%. The reduction in states due to 1-bit parity goes down with the increase in word size because the effect of 1-bit parity gets absorbed by the larger word size. For a 32-bit word the effect of single bit is 1/32 whereas for a 256-bit word the effect reduces to 1/256. The reduction in states is maximum for 16-/bit word, 70% reduction. Below 16-bit the reduction rates drops because there are not many states available to merge. For a 4-bit word there are 5 states out of which 2 are merged by 1-bit parity. Therefore, the 1-bit parity mitigation technique works best for 16-32 bit word sizes.

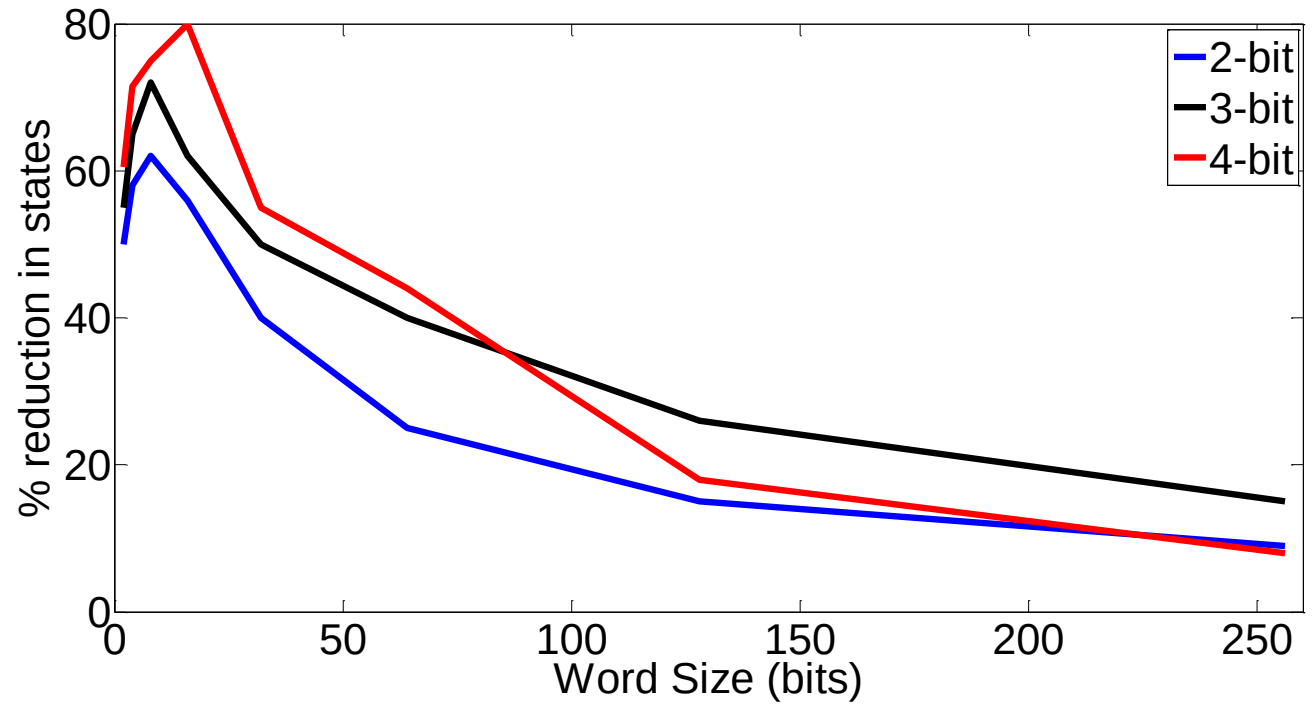

Fig. 8 Percent reduction in states with multi-bit random write for different word sizes.

Note that the overhead associated with parity is negligible for practical word sizes. Furthermore, parity encoding is typically present in the error correction code (ECC) protected memory arrays. Therefore, this technique is easily introduced in the design by reusing existing design features.

### *C. Adding Random bits in Word*

The reduction is states with 1-bit parity diminishes as the word size increases. The signature of the current waveform at higher word sizes becomes difficult to interpret as the number of states increase. To further obfuscate the signature we propose to add multiple random bits in the word during write. This technique further complicates and merges the states in the supply current signature. The results with addition of 2, 3 and 4 random bits in the word is shows in Fig. 8. It can be observed that the larger number of extra random bits reduce the number states substantially for larger word sizes. The random bits can be generated by employing a simple pseudo random number generator. For larger word sizes the overhead from few extra bits is expected to be negligible.

### *D. Constant Current Write*

In the previous section it has been noted that asymmetric polarity dependent write current is a manifestation of constant voltage write. If we write both ‘1’ and ‘0’ with the same amount of current, then there will be only one level in the current waveform and the write current will only depend on the word size. Constant current write can be achieved by using a current mirror with voltage controlled current source (Fig. 9(a)). The two PMOS forms the current mirror whereas the NMOS $M_C$ controls the current to be mirrored depending on the STTRAM resistance [9]. Bias voltage ($V_B$) is adjusted to provide the initial read current in the main branch which will pass through the STTRAM in the auxiliary branch. However constant current write will create mismatch in switching times between ‘0’ and ‘1’ states (Fig. 9(b)). This will affect the design of the word-line driver but the adversary will have no clue about the data as the current will remain constant throughout the write access. Since the difference in switching current between ‘0’ and ‘1’ is ~0.4ns it is challenging for the attacker to extract the entire word using a single supply current signature.

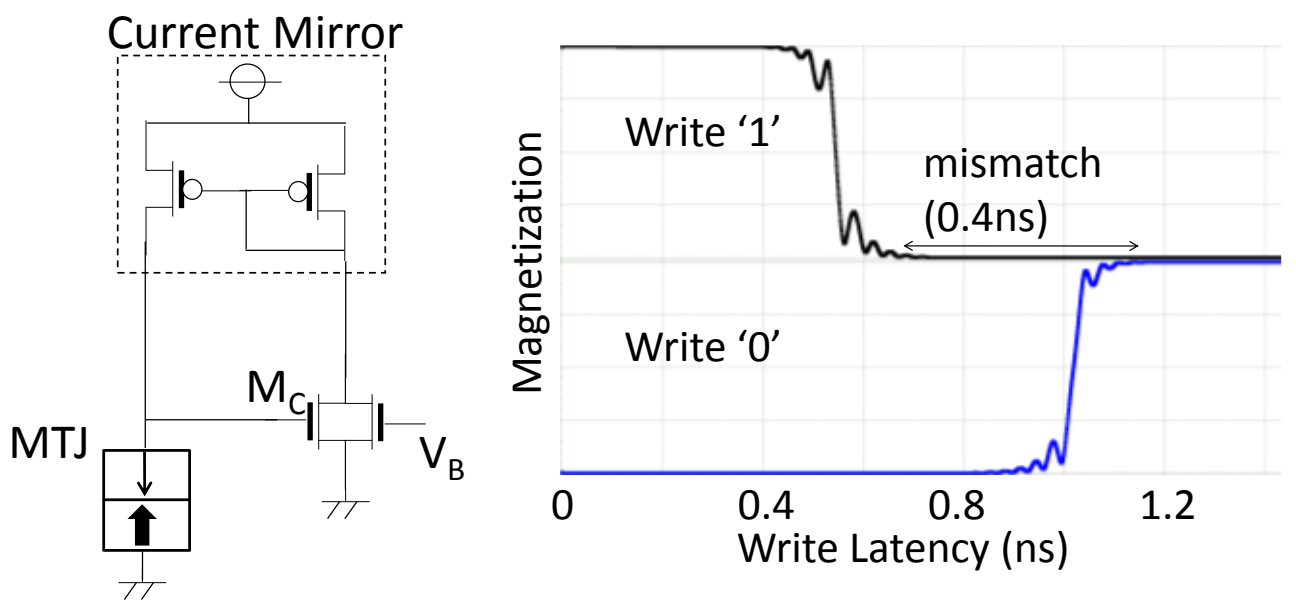

(a) (b)

Fig. 9(a) Constant current write circuit; and, [9] (b) write latency difference with constant current write.

### *E. Word Size*

The supply current waveform highly depends on the number of bits that is being read and written at once i.e., the word size. With the increase in word size and under PV the attack window for the adversary will reduce. This will affect the prediction accuracy and increase the difficulty for the adversary to correctly predict the number of 0’s and 1’s stored in the memory array. Thus, increasing word size during read and write can potentially lower the attack window for the adversary.

## V. Discussions

In this section we discuss the applicability of the proposed attack model and countermeasures for various scenarios.

### *A. Impact of Scaling*

With technology scaling the MTJ size reduces which lowers the free layer thickness. The thermal stability ($\Delta_t$) is linearly dependent on the free layer thickness and the retention time is exponentially related to $\Delta_t$. Therefore the write latency and write current of STTRAM is expected to scale down making it more secure against power analysis attack. Introduction of perpendicular magnetic anisotropy (PMA) STTRAM makes it further challenging for the adversary to perform meaningful side channel attack due to inherently lower write latency and write current offered by this technology.

### *B. Impact of Usage*

Although STTRAM LLC is considered in this paper the proposed attack models are equally applicable to the STTRAM main memory. Availability of dedicated power supply makes it easy to probe main memory active current.

However, cryptographic keys cannot be revealed since the crypto operations are performed on chip. Nevertheless, the raw unencrypted sensitive data can be extracted.

*C. Impact of Magnetic Tampering*

External DC magnetic field of opposite strength could be used to increase the switching time of MTJ which will increase the attack window for the adversary. Thus, with the help of a common horseshoe magnet the adversary can increase the write latency to facilitate the attack (especially for constant voltage write scheme).

*D. Cache Timing Attack*

In shared computer the main memory and hard disk are protected against use by another user on the same machine but the cache is not. If two users are working on the same machine the malicious user can fill the entire cache with his own data and wait for the other user to perform secret operations like encryption. The malicious user then measures the loading time to find which of his data has been replaced by the other user and learns about the cache addresses used in encryption. This timing information can be exploited for key recovery of encryption algorithms like AES [14]. Since a larger cache size can be afforded with STTRAM (due to smaller footprint bitcell) the number of cache line replacements is expected to be less alleviating the cache timing attack. However the persistence of data can be exploited to launch the attack at a later time to retrieve the sensitive information.

*E. Other Side Channels*

STTRAM resistance in the parallel and anti-parallel state is in the range of KΩ (5K-10K) and the write current is in the order of μA (100-150 μA). Thus, the IR drop will be in the order of mV resulting in considerable droop in supply voltage. The adversary can monitor the droops in supply voltage to identify write operation and the amount of droop can give out the information about the data being written much similar to supply current.

## VI. Conclusions

In this paper we showed that STTRAM read/write current, latency and asymmetricity can be security vulnerabilities. We presented novel side channel attack models for STTRAM to compromise the sensitive data in LLC. We also provided a suite of preventive countermeasures such as constant current write, increased word size, SNVM and parity bit encoding to increase the reverse engineering effort required by the adversary to decipher the data from read and write current waveforms. The proposed techniques showed significant promise to protect against data privacy attacks to enable secure NVM design.

## VII. References

[1] Ohsawa, T., H. Koike, S. Miura, H. Honjo, K. Tokutome, S. Ikeda, T. Hanyu, H. Ohno, and T. Endoh. "1Mb 4T-2MTJ nonvolatile STT-RAM for embedded memories using 32b fine-grained power gating technique with 1.0 ns/200ps wake-up/power-off times." In VLSI Circuits (VLSIC), 2012 Symposium on, pp. 46-47. IEEE, 2012.

[2] Jang, Jae-Won, Jongsun Park, Swaroop Ghosh, and Swarup Bhunia. "Self-correcting STTRAM under magnetic field attacks." In Design Automation Conference (DAC), 2015 52nd ACM/EDAC/IEEE, pp. 1-6. IEEE, 2015.

[3] Bi, Xiuyuan, Hai Li, and Jae-Joon Kim. "Analysis and optimization of thermal effect on STT-RAM Based 3-D stacked cache design." In VLSI (ISVLSI), 2012 IEEE Computer Society Annual Symposium on, pp. 374-379. IEEE, 2012.

[4] Rathi, Nitin, Swaroop Ghosh, Anirudh Iyengar and Helia Naeimi. “Data Privacy in Non-Volatile Cache: Challenges, Attack Models and Solutions”, In 2016 Asia and South Pacific Design Automation Conference, IEEE, 2016.

[5] Jameco Electronics, “PC-Multiscope (part# 142834),” p.103, 1999.

[6] Motaman, Seyedhamidreza, Swaroop Ghosh, and Nitin Rathi. "Impact of process-variations in STTRAM and adaptive boosting for robustness." InProceedings of the 2015 Design, Automation & Test in Europe Conference & Exhibition, pp. 1431-1436. EDA Consortium, 2015.

[7] Kim, Jisu, Kyungho Ryu, Seung H. Kang, and Seong-Ook Jung. "A novel sensing circuit for deep submicron spin transfer torque MRAM (STT-MRAM)." Very Large Scale Integration (VLSI) Systems, IEEE Transactions on 20, no. 1 (2012): 181-186.

[8]Swaminathan, Karthik, Raghav Pisolkar, Cong Xu, and Vijaykrishnan Narayanan. "When to forget: A system-level perspective on STT-RAMs." InDesign Automation Conference (ASP-DAC), 2012 17th Asia and South Pacific, pp. 311-316. IEEE, 2012.

[9] Halupka, David. "Effects of Silicon Variation on Nano-Scale Solid-State Memories." PhD diss., University of Toronto, 2011.

[10] Eken, Enes, Yaojun Zhang, Wujie Wen, Rajan Joshi, Huaqing Li, and Yiran Chen. "A Novel Self-Reference Technique for STT-RAM Read and Write Reliability Enhancement." Magnetics, IEEE Transactions on 50, no. 11 (2014): 1-4.

[11] Halderman, J. Alex, Seth D. Schoen, Nadia Heninger, William Clarkson, William Paul, Joseph A. Calandrino, Ariel J. Feldman, Jacob Appelbaum, and Edward W. Felten. "Lest we remember: cold-boot attacks on encryption keys." Communications of the ACM 52, no. 5 (2009): 91-98.

[12] Kocher, Paul, Joshua Jaffe, and Benjamin Jun. "Differential power analysis." In Advances in Cryptology—CRYPTO’99, pp. 388-397. Springer Berlin Heidelberg, 1999.

[13] Smullen, Clinton W., Vidyabhushan Mohan, Anurag Nigam, Sudhanva Gurumurthi, and Mircea R. Stan. "Relaxing non-volatility for fast and energy-efficient STT-RAM caches." In High Performance Computer Architecture (HPCA), 2011 IEEE 17th International Symposium on, pp. 50-61. IEEE, 2011.

[14] Bernstein, Daniel J. "Cache-timing attacks on AES." (2005).